\documentclass[aps,prc,superscriptaddress,nofootinbib,showpacs,floatfix]{revtex4-2}
 
\usepackage{graphicx}
\usepackage{dcolumn}
\usepackage{bm}
\usepackage{amsmath}
\usepackage{hyperref}
\usepackage[mathlines]{lineno}
\usepackage{hyperref}
\usepackage{xspace}
\usepackage{xcolor}

\newcommand{\sqn}{\sqrt{s_{_{\mathrm{NN}}}}}
\newcommand{\Nch}{\textit{N}_\mathrm{ch}}
\newcommand{\pT}{p_\mathrm{T}}
\newcommand{\npart}{\mbox{$N_{\mathrm{part}}$}}

\begin{document}
\title{A Method to Constrain Preferential Emission and Spectator Dynamics in Heavy-Ion Collisions}

\author{Vipul Bairathi}%
\affiliation{%
 Instituto de Alta Investigación, Universidad de Tarapacá\\
 Casilla 7D, Arica 1000000, Chile
}

\author{Somadutta Bhatta}
\email{s.bhatta@uu.nl}
 \affiliation{
 Institute for Gravitational and Subatomic Physics (GRASP), Utrecht University,\\
 Princetonplein 1, 3584 CC Utrecht, Netherland
}
\affiliation{Nikhef, Science Park 105, 1098 XG Amsterdam, The Netherlands}

\date{\today}

\begin{abstract}
Longitudinal particle production in heavy-ion collisions is influenced both by preferential emission from participating nucleons and by the breakup of spectator matter, yet quantifying these effects experimentally remains challenging. We introduce a Pearson correlation between spectator and charged-particle forward-backward asymmetries as an experimental probe of these phenomena. Using Au+Au collisions at $\sqrt{s_{NN}}=200$~GeV simulated with A Multi-Phase Transport (AMPT) model, we validate that this correlator provides a robust, pseudorapidity-differential measure of the influence of preferential emission on the longitudinal structure of particle production. We further demonstrate that the correlation strength is sensitive to fluctuations in spectator number, which in experiments arise from evaporation and fragmentation of the spectator remnants. The proposed observable therefore offers a data-driven handle for constraining models of preferential emission and spectator breakup, thereby improving our understanding of the mechanisms that shape the final-state longitudinal distributions in heavy-ion collisions.
\end{abstract}


\maketitle

\section{Introduction}\label{sec:Introduction}
In heavy-ion collisions, the initial-state geometry and the subsequent evolution of the produced Quark-Gluon Plasma (QGP) leave distinct imprints on the azimuthal and longitudinal distributions of final-state particles. Over the past two decades, extensive studies of azimuthal anisotropies in the final state have yielded stringent constraints on the initial-state geometry and the transport properties of the QGP. In contrast, the dynamics governing the pseudorapidity ($\eta$) dependence of particle production are relatively less explored. A broad set of measurements and phenomenological studies such as forward–backward multiplicity correlations, flow decorrelations, and baryon stopping—have advanced our understanding of the three-dimensional energy-density profile of the initial state and its fluctuations~\cite{Jia:2014vja, Jia:2017kdq, ATLAS:2017rij, ATLAS:2020sgl, Jia:2015jga, ATLAS:2012as, Gazdzicki:2005rr}. 

The longitudinal distribution of the produced particles, $P(\eta)$, is shaped by two contributions: (i) \emph{preferential emission}, which primarily affects mid and intermediate pseudorapidities, and (ii) \emph{spectator fragmentation}, which influences $P(\eta)$ near the beam rapidity, $y_{\mathrm{beam}}$. Preferential emission has been explored through its manifestations in long-range multiplicity correlations~\cite{Jia:2015jga}, longitudinal flow decorrelations, and directed flow~\cite{Jia:2014vja, Bozek:2010bi, Bozek:2015bna, Pang:2015zrq, Rohrmoser:2019xis}. In contrast, the space-time evolution of spectator dynamics and their impact on $P(\eta)$ remain far less constrained~\cite{Shen:2022oyg, Du:2023efk}.

Preferential emission is typically described using a nucleon fragmentation function, which governs the $\eta$-distribution of hadrons produced by each participating nucleon and peaks in the direction of that nucleon’s initial motion. Event-by-event fluctuations in the numbers of forward- and backward-going participants, $N_{\mathrm{part,F}}$ and $N_{\mathrm{part,B}}$, therefore generate forward-backward asymmetries in $P(\eta)$. Because each participant nucleon emits more hadrons along its original direction~\cite{Bialas:2004su, Bzdak:2009xq, Bzdak:2009dr}, a difference $N_{\mathrm{part,F}} - N_{\mathrm{part,B}}$ induces an $\eta$-odd structure in the final-state pseudorapidity distribution of produced charged particles $P(\eta)$. This mechanism underlies observed features of long-range multiplicity correlations, flow decorrelations, and directed flow.

Spectator fragmentation constitutes the second major contribution to the longitudinal structure, particularly at lower collision energies or in more peripheral events~\cite{PHOBOS:2010eyu}. In the Abrasion-Ablation picture~\cite{Gaimard:1991vb, Hufner:1975zz}, the initial \textit{abrasion} stage removes participant nucleons from the colliding nuclei, leaving the spectator remnants in an excited state; these remnants subsequently \textit{ablate} through evaporation and fragmentation. In heavy-ion collisions, the resulting spectator fragments can deposit charged clusters over a broad pseudorapidity range centered near $y_{\mathrm{beam}}$~\cite{PHOBOS:2015aeh, Tarafdar:2014oua}. Such contributions can significantly alter forward-rapidity multiplicities and must therefore be quantified when interpreting longitudinal observables. Taken together, these considerations motivate the development of an experimentally accessible collider observable that directly correlates spectator asymmetry with the pseudorapidity-odd structure of the produced-particle distribution.

The relationship between participant motion and particle production was categorized in the fixed-target study in Ref.~\cite{Gazdzicki:2005rr}. Using NA49 Pb+Pb data at $\sqrt{s_{_{\mathrm{NN}}}}=17.3~\mathrm{GeV}$, this study examined multiplicity fluctuations in event classes of fixed projectile participants to distinguish among three limiting scenarios of longitudinal dynamics:  (i) \emph{transparency}, where projectile and target matter retain their initial longitudinal motion;  (ii) \emph{mixing}, where substantial longitudinal stopping causes the sources to overlap and produce particles closer to midrapidity; and  (iii) \emph{reflection}, where projectile (target) matter is deflected into the backward (forward) hemisphere. The NA49 results favored significant mixing over the transparency or reflection limits. Crucially, this analysis demonstrated experimentally that correlations between spectator numbers and produced-particle yields provide direct information about the degree of longitudinal stopping.

A key question is how these longitudinal dynamics evolve with increasing collision energy—potentially transitioning from mixing toward greater transparency at RHIC and LHC energies. However, differential observables that directly quantify how initial longitudinal fluctuations translate into the pseudorapidity dependence of produced particles remain limited. Existing measurements typically rely on comparisons with event generators. In addition, existing measurements provide only limited constraints on current models attempting to describe the complex breakup of spectator matter~\cite{PHOBOS:2015aeh, STAR:2014clz}.

Motivated by the principle that correlations between the number of participants and spectators can probe longitudinal stopping, we construct an experimentally accessible collider observable that correlates the event-by-event forward–backward asymmetry of spectator matter with the pseudorapidity-odd component of final-state charged-particle production. Formulated in terms of experimentally accessible quantities, the method enables a systematic investigation of how source asymmetries are imprinted on $P(\eta)$ and provides a means to quantify the contribution of spectator fragmentation. In doing so, it extends the insights gained from fixed-target SPS-era measurements of preferential emission to the RHIC and LHC energy regimes, while simultaneously placing new constraints on spectator dynamics.

This paper is structured as follows. Section~\ref{sec:Methodology} first details the methodology used to derive the proposed correlator. Then, Section~\ref{sec:Results} presents the primary findings of this study, reporting on the sensitivity of the correlator to preferential emission and spectator fragmentation. Finally, Section~\ref{sec:conclusion} provides the conclusion and outlook.

\section{Methodology}\label{sec:Methodology}
We use the AMPT model to study Au+Au collisions at $\sqn=200$~GeV, as it successfully reproduces a wide range of experimental observables~\cite{Lin:2004en, Jia:2015jga, Jia:2016jlg, Jia:2022qgl, Xu:2011jm, Xu:2011fe, Solanki:2012ne, Jia:2022qrq, Nielsen:2022jms, Magdy:2020gxf, Lin:2021mdn}. In our analysis, the final-state charged-particle multiplicity, $\Nch$, includes pions, kaons, and protons with $\pT<10$~GeV. Centrality is determined using the charged-particle multiplicity within $|\eta|<0.2$. To avoid autocorrelations, all observables are calculated using particles with $|\eta|>0.2$. A systematic check, where centrality is defined using $\Nch$ within $|\eta|<0.5$, yields results consistent with those obtained using $|\eta|<0.2$, indicating no impact on the results or their interpretation.

In the model context, the total number of participants is $\npart = N_{\mathrm{part,F}} + N_{\mathrm{part,B}}$, where $N_{\mathrm{part,F}}$ and $N_{\mathrm{part,B}}$ denote the number of forward- and backward-going nucleons, respectively. Similarly, $N_{\mathrm{ch},\eta}$ and $N_{\mathrm{ch},-\eta}$ represent the charged-particle multiplicities in narrow symmetric intervals, $\delta\eta$ around $\pm\eta$. The participant asymmetry, $a_{\mathrm{part}}$, and the final-state charged-particle asymmetry, $a_{\mathrm{ch},\eta}$, are defined as:
\begin{align}
\label{eq:basic_defs}
a_{\mathrm{ch},\eta} & = N_{\mathrm{ch},\eta} - N_{\mathrm{ch},-\eta}, \\
a_{\mathrm{part}}    & = N_{\mathrm{part,F}} - N_{\mathrm{part,B}}.
\end{align}

The total number of spectators in each event is  $N_{\mathrm{spec}} = 2A - \npart$, where $A$ is the nuclear mass number. The spectator asymmetry is  $a_{\mathrm{sp}} = N_{\mathrm{spec,F}} - N_{\mathrm{spec,B}}$,  and since $N_{\mathrm{part}} + N_{\mathrm{spec}} = 2A$ is fixed, the participant and spectator asymmetries are related by  $a_{\mathrm{part}} = -a_{\mathrm{sp}}$. Correlating $a_{\mathrm{ch},\eta}$ with $a_{\mathrm{sp}}$ thus quantifies the relationship between the initial-state participant asymmetry and the final-state asymmetries in $P(\eta)$. To reduce detector effects, we use normalized variables for both asymmetries:
\begin{equation}
\label{eq:norm_asym}
\alpha_{\mathrm{sp}} = 
\frac{N_{\mathrm{spec,F}} - N_{\mathrm{spec,B}}}
     {N_{\mathrm{spec,F}} + N_{\mathrm{spec,B}}},
\qquad
\alpha_{\mathrm{ch},\eta} = 
\frac{N_{\mathrm{ch},\eta} - N_{\mathrm{ch},-\eta}}
     {N_{\mathrm{ch},\eta} + N_{\mathrm{ch},-\eta}}.
\end{equation}
These ratios cancel the proportionality factors arising from detector inefficiencies to first order, enabling a robust measurement across a wide $\eta$ range. We then correlate the two normalized quantities using a Pearson correlation coefficient.

To relate these variables to experimentally measurable quantities, we note that $\alpha_{\mathrm{ch},\eta}$ is readily constructed from multiplicity measurements provided by tracking detectors or calorimeters covering mid- and forward pseudorapidities. Conversely, the experimental estimation of $\alpha_{\mathrm{sp}}$ relies on Zero Degree Calorimeters (ZDCs).  ZDCs are compact hadronic calorimeters located at very forward angles ($\theta \approx 0^\circ$) relative to the beam line, typically situated downstream of dipole magnets. Because these magnets deflect charged spectator fragments (protons) away from the beam axis, ZDCs are primarily sensitive to spectator neutrons. The number of these neutrons is determined by measuring the total energy deposited in the forward and backward ZDCs and dividing by the beam energy per nucleon. Thus, in experimental applications, only the asymmetry of spectator neutrons can be measured. The effect of using neutron spectator asymmetry versus total spectator asymmetry is estimated within the model study later in this paper.

Calculating the Pearson correlator requires the variances of $\alpha_{\mathrm{sp}}$ and $\alpha_{\mathrm{ch},\eta}$. While $\alpha_{\mathrm{sp}}$ is determined by the global spectator configuration, the magnitude of $\alpha_{\mathrm{ch},\eta}$ depends on the pseudorapidity bin width, $\delta\eta$. To eliminate this dependence, we introduce a modified variance. A first-order Taylor expansion shows that for symmetric collisions, $\mathrm{Var}(\alpha_{\mathrm{ch},\eta}) \approx \sigma^2(N_{\mathrm{ch},\eta})/(2\mu^2(N_{\mathrm{ch},\eta}))$. Assuming a locally uniform particle density, both the mean and variance of $N_{\mathrm{ch},\eta}$ scale linearly with $\delta\eta$, implying that $\mathrm{Var}(\alpha_{\mathrm{ch},\eta})$ scales as $1/\delta\eta$ (see Appendix~\ref{Sec:Appendix1}). We therefore define the modified variance as $\mathrm{Var}(\alpha_{\mathrm{ch},\eta})_{\mathrm{mod}} = \mathrm{Var}(\alpha_{\mathrm{ch},\eta}) \cdot \delta\eta$. This ensures the correlator is invariant under changes in $\delta\eta$. The final correlator is defined as,
\begin{equation}
\label{eq:rhodecomp}
\rho(\alpha_{\mathrm{sp}},\alpha_{\mathrm{ch},\eta})
= -\,\frac{\mathrm{Cov}(\alpha_{\mathrm{sp}},\alpha_{\mathrm{ch},\eta})}
{\sqrt{\mathrm{Var}(\alpha_{\mathrm{sp}})\,
       \mathrm{Var}(\alpha_{\mathrm{ch},\eta})_{\mathrm{mod}}}},
\end{equation}
where the minus sign accounts for the inverse relation between spectator and participant asymmetries, $\alpha_{\mathrm{sp}} \sim -\alpha_{\mathrm{part}}$. In symmetric collision systems, the means $\langle \alpha_{\mathrm{sp}} \rangle$ and  $\langle \alpha_{\mathrm{ch},\eta} \rangle$  are approximately zero, so the covariance simplifies to  $\langle \alpha_{\mathrm{sp}}\,\alpha_{\mathrm{ch},\eta}\rangle$ and the variance to  $\langle (\alpha_{\mathrm{sp}})^2\rangle$.

It is broadly understood that spectators are space-time separated from the hot midrapidity medium~\cite{PHOBOS:2015aeh}, and that forward/backward spectator fragmentation proceeds largely independently of midrapidity particle production~\cite{Tarafdar:2014oua}. Accordingly, when $\alpha_{\mathrm{ch},\eta}$ is constructed from midrapidity particles with a sufficient $\eta$-gap to the spectator acceptance, spectator dynamics are not expected to contribute to $\mathrm{Cov}\!\left(\alpha_{\mathrm{sp}},\alpha_{\mathrm{ch},\eta}\right)$. For very forward $|\eta|$ close to beam rapidity, spectator fragments can populate the $\alpha_{\mathrm{ch},\eta}$ windows, yielding a nonzero covariance. In our AMPT-based study, this independence is automatic since spectator fragments are not explicitly transported into the midrapidity region.

However, the denominator of $\rho$ contains $\mathrm{Var}\!\left(\alpha_{\mathrm{sp}}\right)$, which is enhanced by fluctuations in the number of detected spectators arising from fragmentation/evaporation and intrinsic Fermi motion \cite{Tarafdar:2014oua}. An increased $\mathrm{Var}\!\left(\alpha_{\mathrm{sp}}\right)$ reduces $|\rho|$. Thus, suppression of  $\rho\!\left(\alpha_{\mathrm{sp}},\alpha_{\mathrm{ch},\eta}\right)$ in experimental data relative to models that neglect spectator breakup provides a data-driven constraint on net broadening of spectator-number distribution due to increased fluctuations arising from spectator matter fragmentation and evaporation.

Before presenting the results, we correct a known issue in the publicly available AMPT version (v2.26t9b), which produces an unphysical forward-backward asymmetry in $dN_{\mathrm{ch}}/d\eta$ for symmetric Au+Au collisions. Because the correlator $\rho(\alpha_{\mathrm{sp}},\alpha_{\mathrm{ch},\eta})$ is explicitly sensitive to pseudorapidity-odd structure, this artifact must be removed. Following the procedure detailed in Appendix~\ref{Sec:Appendix2}, we apply an $\eta$-dependent weight in each centrality class on an event-by-event basis to restore the expected symmetry $dN/d\eta(\eta)=dN/d\eta(-\eta)$ in each centrality, while preserving the overall yield. All the results shown in this work use the symmetrized sample from the AMPT model.

\section{Results}\label{sec:Results}
\begin{figure*}[!htbp]
\centering
\includegraphics[width=0.32\textwidth]{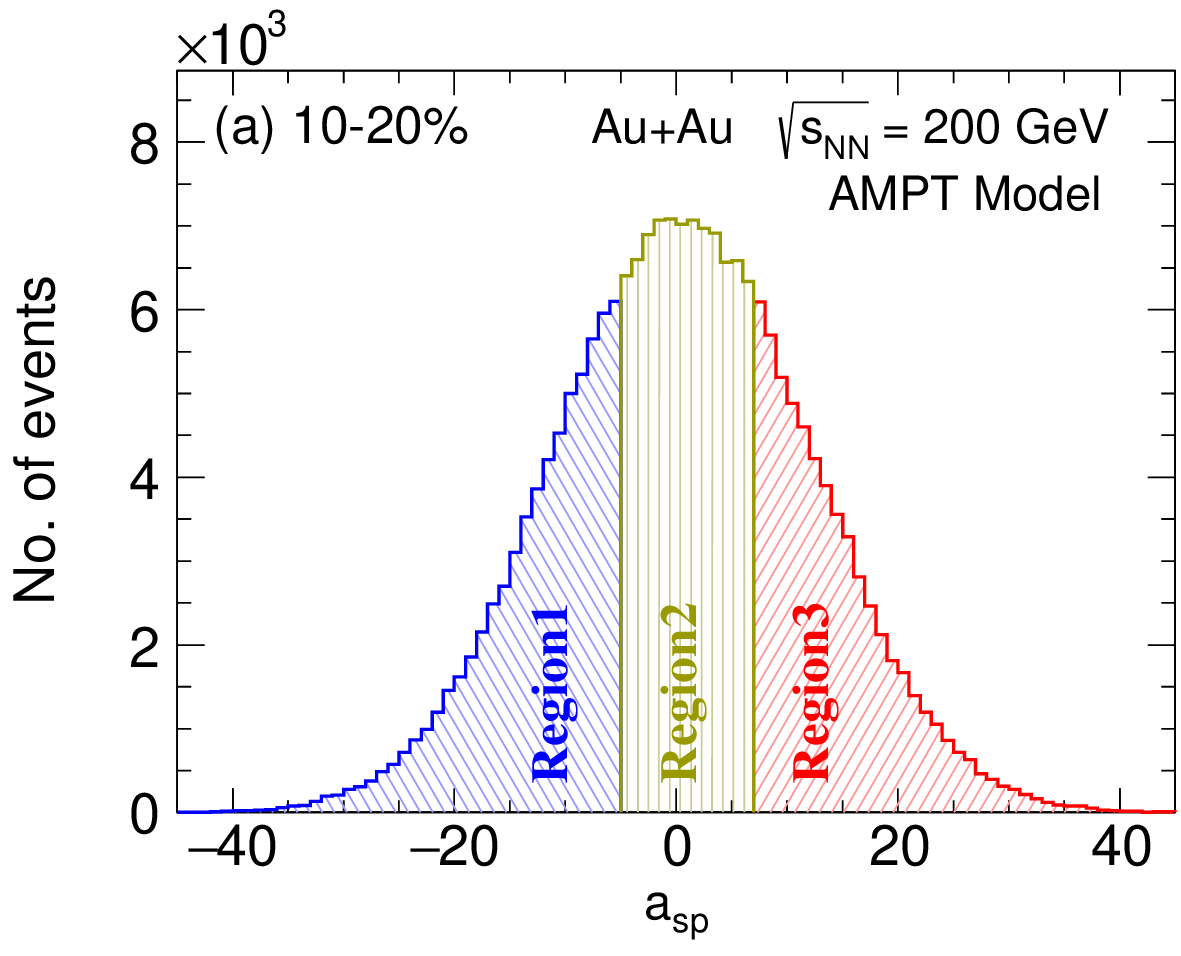}
\includegraphics[width=0.32\textwidth]{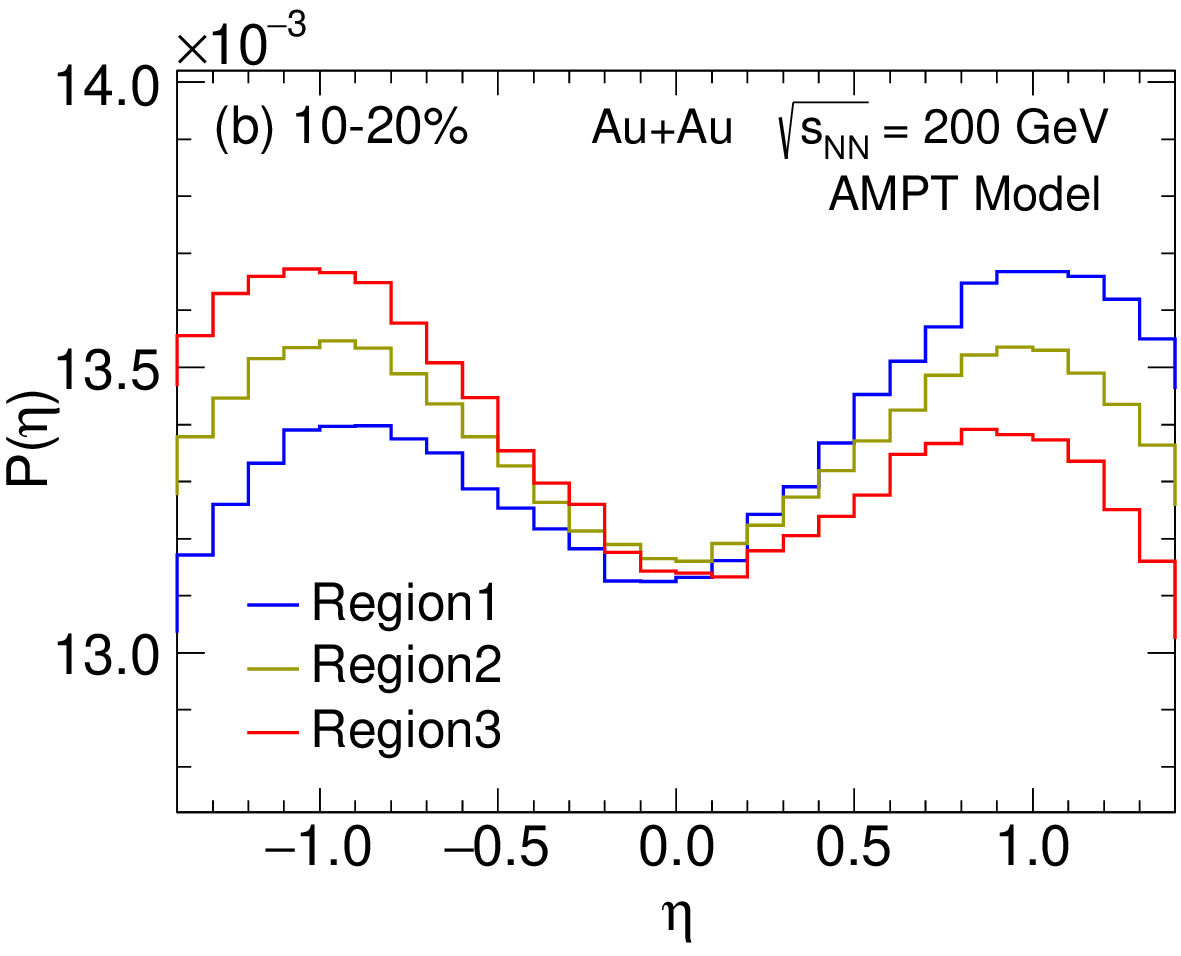}
\includegraphics[width=0.32\textwidth]{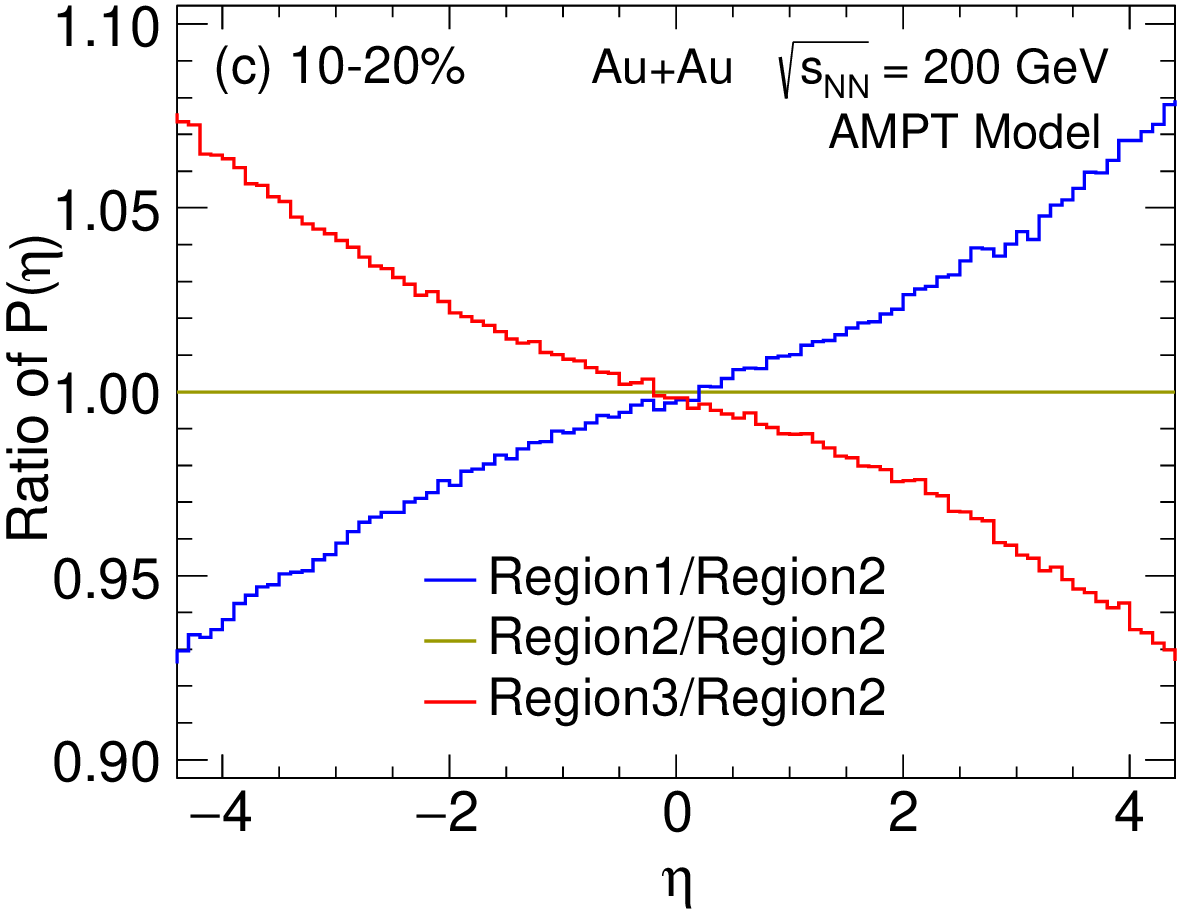}
\caption{(a) Distribution of $a_{\mathrm{sp}}$ in 10-20\% central Au+Au collisions at $\sqn=200$~GeV. Regions 1, 2, and 3 denote $a_{\mathrm{sp}}\ll 0$, $a_{\mathrm{sp}}\approx 0$, and $a_{\mathrm{sp}}\gg 0$, respectively. (b) Comparison of the $P(\eta)$ for events in each region in panel a. (c) Ratios of the $P(\eta)$ for events with asymmetric participant configurations (Regions 1 and 3) to those with symmetric configurations (Region 2).}
\label{fig:main1}
\end{figure*}

We first demonstrate how preferential emission influences $P(\eta)$. Figure~\ref{fig:main1}(a) shows the distribution of spectator asymmetry, $a_{\mathrm{sp}}$, for 10-20\% central Au+Au collisions $\sqn=200$~GeV from the AMPT model. We categorize the events into three groups based on this asymmetry: Region 1 ($a_{\mathrm{sp}} \ll 0$, indicating more forward-moving participants), Region 2 ($a_{\mathrm{sp}} \approx 0$, indicating a symmetric distribution), and Region 3 ($a_{\mathrm{sp}} \gg 0$, indicating more backward-moving participants). As expected from preferential emission, events characterized by more forward-moving participants exhibit an increased particle yield at positive $\eta$, as shown in Fig.~\ref{fig:main1}(b). The ratios of the particle distributions depicted in Fig.~\ref{fig:main1}(c) highlight this effect, confirming that the initial-state participant asymmetry directly translates to final-state pseudorapidity asymmetry in the AMPT model~\cite{ZiWei2022RBRC}.

\begin{figure}[!htbp]
\centering
\includegraphics[width=0.35\textwidth]{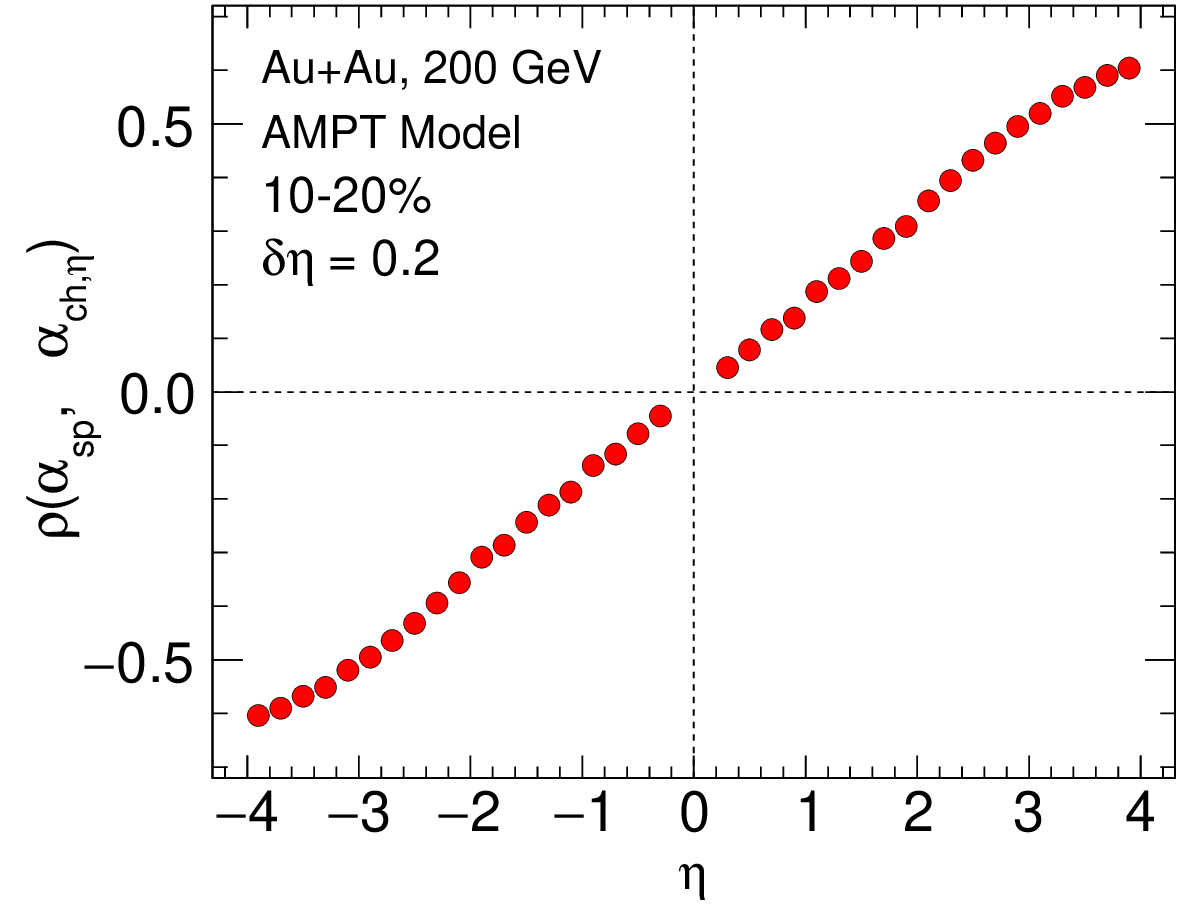}
\caption{The correlation coefficient $\rho(\alpha_{\mathrm{sp}},\alpha_{\mathrm{ch},\eta})$ for $\delta\eta=0.2$ in 10--20\% central Au+Au collisions at $\sqn=200$~GeV from the AMPT model, exhibiting an odd dependence on $\eta$ characteristic of preferential emission.}
\label{fig:main2}
\end{figure}

\begin{figure}[!htbp]
\centering
\includegraphics[width=0.7\textwidth]{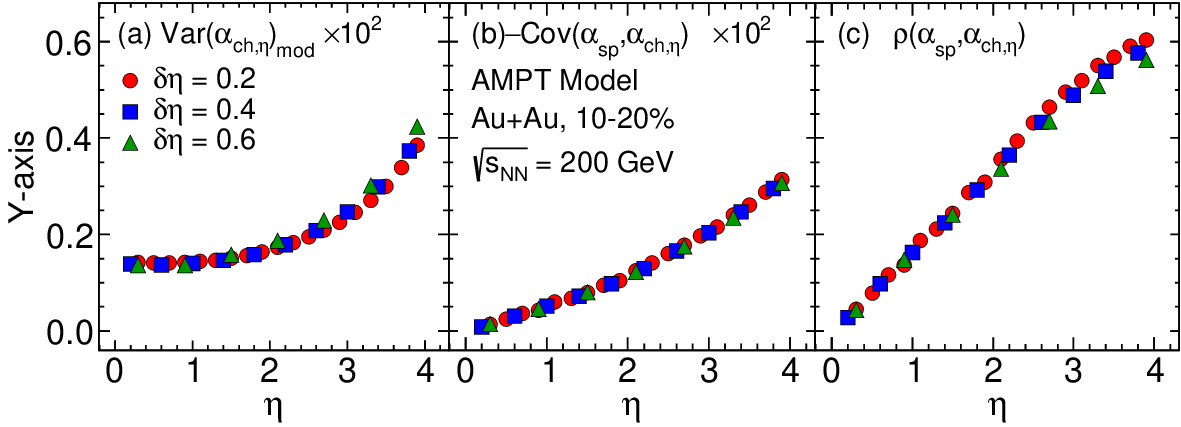}
\caption{(a) $\mathrm{Var}(\alpha_{\mathrm{ch},\eta})_{\mathrm{mod}}$, (b) $-\mathrm{Cov}(\alpha_\mathrm{sp},\alpha_{\mathrm{ch},\eta})$, and (c) $\rho(\alpha_\mathrm{sp},\alpha_{\mathrm{ch},\eta})$ as a function of $\eta$ for different bin-widths $\delta\eta$ in 10-20\% central Au+Au collisions at $\sqn = 200$ GeV from the AMPT model. The quantities are arbitrarily scaled for better visibility.}
\label{fig:method_robustness}
\end{figure}

To quantify this relationship differentially, we calculate the correlator $\rho(\alpha_{\mathrm{sp}},\alpha_{\mathrm{ch},\eta})$, as shown in Fig.~\ref{fig:main2}. The correlation is clearly an odd function of $\eta$, vanishing at $\eta = 0$ and increasing in magnitude towards forward $\eta$. This observation demonstrates that particles produced at larger $|\eta|$ retain a stronger ``memory" of the direction of their parent nucleons.
As a cross-check, the robustness of the proposed method is demonstrated in Fig.~\ref{fig:method_robustness}. The components of the correlator, which include the modified variance, covariance, and the $\rho$ itself, are shown for several choices of $\delta\eta$. As expected from the derivation, all three quantities remain largely unchanged, confirming that the observable is insensitive to variations in either the width of the $\eta$ window or the measurement granularity. 

In deriving $\mathrm{Var}(\alpha_{\mathrm{ch},\eta})_{\mathrm{mod}}$ in Appendix~\ref{Sec:Appendix1}, we assumed that the pseudorapidity density $P(\eta)$ is approximately constant within each narrow $\eta$ bin. This assumption is violated in regions where $P(\eta)$ changes rapidly, such as at forward $\eta$. In those regions, this violation leads to small residual deviations between different choices of $\delta\eta$ for $\mathrm{Var}(\alpha_{\mathrm{ch},\eta})_{\mathrm{mod}}$ in Fig.~\ref{fig:method_robustness}.

\begin{figure}[!htbp]
\centering
\includegraphics[width=0.35\textwidth]{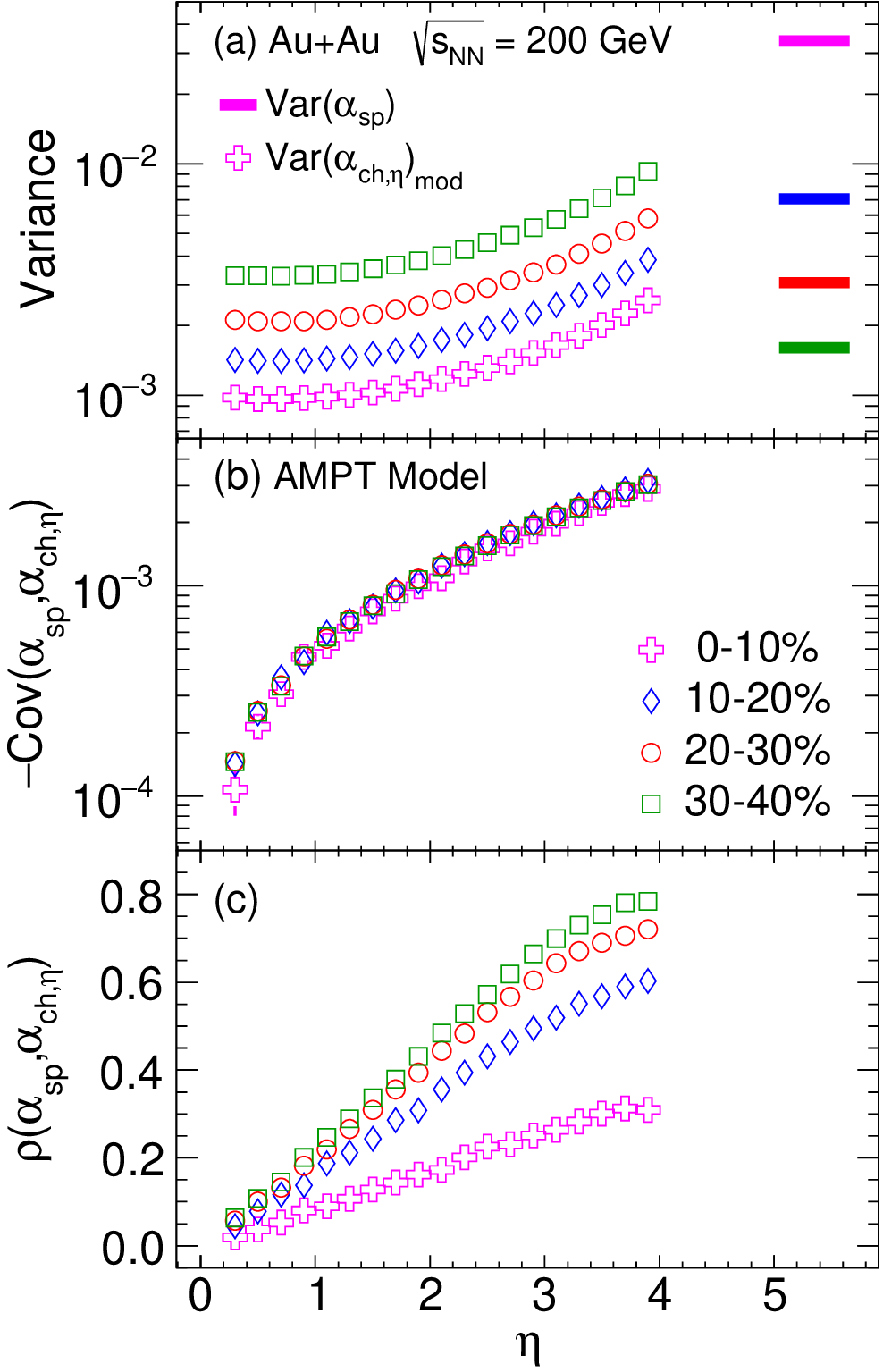}
\caption{The components of the correlator for different centrality intervals in Au+Au collisions at $\sqn=200$~GeV from the AMPT model: (a) modified variance of charged-particle asymmetry, $\mathrm{Var}(\alpha_{\mathrm{ch},\eta})_{\mathrm{mod}}$. (b) covariance, $-\,\mathrm{Cov}(\alpha_{\mathrm{sp}},\alpha_{\mathrm{ch},\eta})$. (c) correlation coefficient, $\rho(\alpha_{\mathrm{sp}},\alpha_{\mathrm{ch},\eta})$. The horizontal lines in (a) show $\mathrm{Var}(\alpha_{\mathrm{sp}})$ for each centrality.}
\label{fig:main3}
\end{figure}

Figure~\ref{fig:main3} illustrates the centrality and $\eta$ dependence of the components of $\rho(\alpha_\mathrm{sp},\alpha_{\mathrm{ch},\eta})$. Within the AMPT model, which does not include spectator fragmentation, $\mathrm{Var}(\alpha_{\mathrm{sp}})$ is independent of $\eta$ and is therefore represented as horizontal lines at $y_{\mathrm{beam}}$ for $\sqrt{s_{NN}} = 200$ GeV. Notably, $\mathrm{Var}(\alpha_{\mathrm{sp}})$ increases toward central collisions. In these events, fewer spectators lead to a smaller denominator in $\alpha_{\mathrm{sp}}$ (Eq.~\ref{eq:norm_asym}), which amplifies the magnitude of fluctuations in the normalized forward-backward spectator asymmetry. Since $\alpha_{\mathrm{part}} \propto -\alpha_{\mathrm{sp}}$, measuring $\mathrm{Var}(\alpha_{\mathrm{sp}})$ also provides information about the initial-state forward-backward participant asymmetry.

In contrast, $\mathrm{Var}\left(\alpha_{\mathrm{ch},\eta}\right)_{\mathrm{mod}}$ decreases from peripheral to central collisions, reflecting a more symmetric $\eta$ distribution in central events. This behavior is expected due to increasingly $\eta$-symmetric particle production sources (strings in the AMPT model), as also reported in Ref.~\cite{Pang:2015zrq}.

The magnitude of $-\mathrm{Cov}(\alpha_{\mathrm{sp}},\alpha_{\mathrm{ch},\eta})$ grows with $|\eta|$ but shows only weak centrality dependence, as shown in Fig.\ref{fig:main3}(b). This indicates that, within the AMPT model, the correlation between participant asymmetry and the $\eta$-asymmetry of produced particles is largely independent of collision geometry. This observation aligns with the expectation that the fragmentation function per participant is centrality-independent\footnote{In AMPT, a string-fragmentation function is used instead of nucleon fragmentation function~\cite{Lin:2004en}}~\cite{Bzdak:2009xq}.

Finally, the $\eta$-dependence of $\rho\left(\alpha_{\mathrm{sp}},\alpha_{\mathrm{ch},\eta}\right)$ in Fig.~\ref{fig:main3}(c) closely follows that of $-\mathrm{Cov}(\alpha_{\mathrm{sp}},\alpha_{\mathrm{ch},\eta})$. Its centrality dependence, however, is driven by the interplay of two variances: $\mathrm{Var}(\alpha_{\mathrm{sp}})$, which captures fluctuations in spectator number, and $\mathrm{Var}\left(\alpha_{\mathrm{ch},\eta}\right)_{\mathrm{mod}}$, which reflects longitudinal asymmetry in the initial particle production sources. Therefore, measuring $\rho\left(\alpha_{\mathrm{sp}},\alpha_{\mathrm{ch},\eta}\right)$ provides a valuable handle to constrain fluctuations in both the number and $\eta$-distribution of particle production sources in heavy-ion collisions.

A crucial aspect of this work is the correlator's sensitivity to the dynamics of spectator fragmentation. The spectators undergo de-excitation through evaporation and fragmentation, which alters the number of free neutrons that reach the ZDCs. To investigate this, we compare the correlator $\rho$ computed using three different definitions of spectators: (1) Total spectators ($N_{\mathrm{spec}}$), the number obtained from the AMPT model, (2) Neutron spectators ($N_{\mathrm{spec}}^{n}$), the neutrons among the spectators as derived from the AMPT model, which does not include effects of spectator fragmentation, and (3) Free-neutron spectators ($N_{\mathrm{spec}}^{fn}$), the number of neutrons detected by the ZDCs in real experiments, where the spectators have undergone fragmentation. The $N_{\mathrm{spec}}^{fn}$ is estimated from measurements by the PHENIX experiment~\cite{Tarafdar:2014oua,PHENIX:2004vdg}. This parameterization accounts for the spectator breakup process. 

\begin{figure}[!htbp]
\centering
\includegraphics[width=0.6\textwidth]{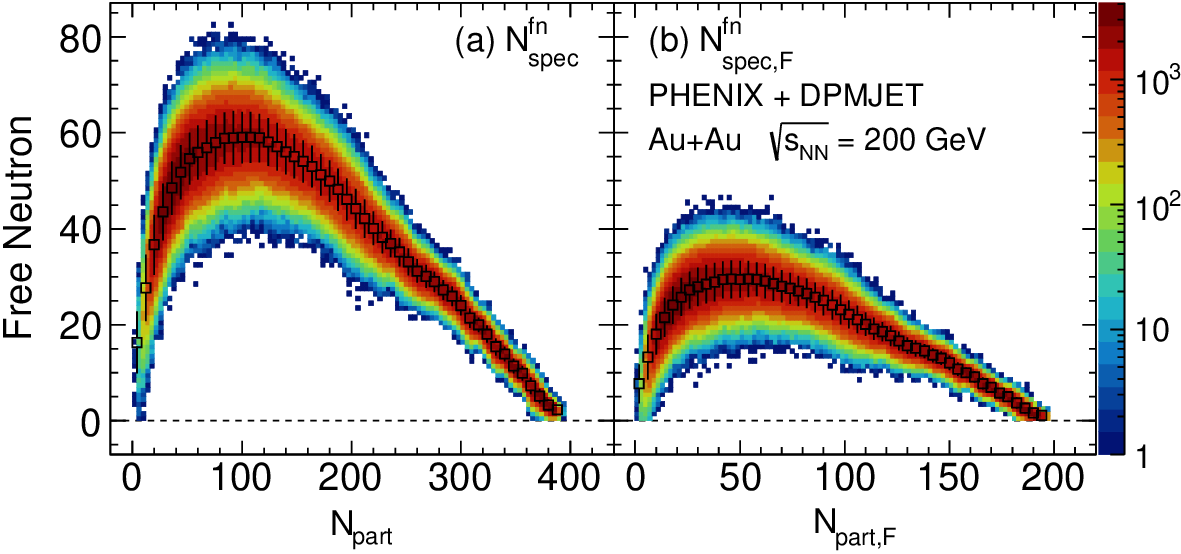}
\caption{(a) The number of free-neutron spectators ($N_{\mathrm{spec}}^{fn}$) versus total participants ($\npart$) for Au+Au collisions at $\sqn = 200$ GeV, simulated using the mean and sigma from the PHENIX measurements~\cite{Tarafdar:2014oua, PHENIX:2004vdg}. (b) The estimated correlation between forward-going free neutrons ($N_{\mathrm{spec,F}}^{fn}$) and forward-going participants ($N_{\mathrm{part,F}}$), which is used to model spectator fragmentation.}
\label{fig:fspec_model}
\end{figure}

The modeling of free-neutron spectators is detailed in Fig.~\ref{fig:fspec_model}. Panel (a) shows the correlation between the number of free neutrons and the total number of participants, using the mean ($\mu$) and width ($\sigma$) derived from measurements by the PHENIX experiment. To apply this to the forward/backward asymmetry, we estimate the relation between the forward-going free neutrons ($N_{\mathrm{spec,F}}^{fn}$) and the forward-going participants ($N_{\mathrm{part,F}}$), as shown in panel (b), following the procedure outlined in Ref.~\cite{Jia:2015jga}.

\begin{figure}[!htbp]
\centering
\includegraphics[width=0.45\textwidth]{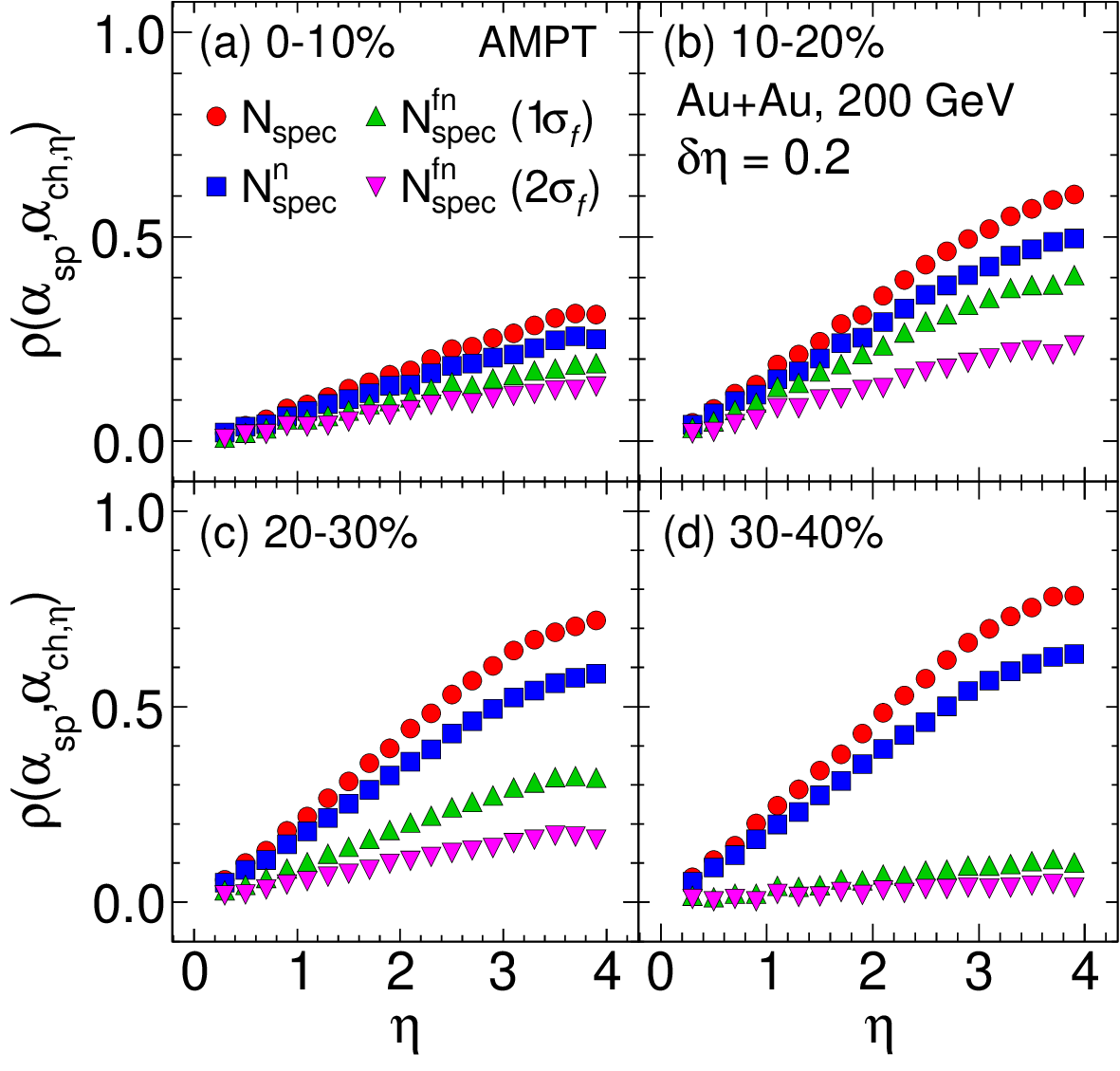}
\caption{Comparison of $\rho(\alpha_{\mathrm{sp}},\alpha_{\mathrm{ch},\eta})$ constructed with total spectators ($N_{\mathrm{spec}}$), neutron spectators without fragmentation ($N_{\mathrm{spec}}^{n}$), and free-neutron spectators with data-driven fragmentation ($N_{\mathrm{spec}}^{fn}$). The strong suppression and centrality dependence of the correlator with $N_{\mathrm{spec}}^{fn}$ highlights its sensitivity to spectator dynamics.}
\label{fig:sensitivity_plot}
\end{figure}

Figure~\ref{fig:sensitivity_plot} compares the correlator for three spectator definitions. Its magnitude is about 20\% smaller for neutron spectators than for total spectators, with an even stronger reduction when spectator fragmentation is taken into account. Using a data-driven estimate of free neutrons ($N_{\mathrm{spec}}^{fn}$), the correlation decreases significantly and exhibits a pronounced centrality dependence: it remains similar to the unfragmented case in central collisions (0-10\%) but nearly vanishes in peripheral collisions (30-40\%) due to increased spectator fragmentation. This centrality dependence is expected because peripheral collisions involve more spectator nucleons, increasing the likelihood of fragmentation. The resulting fluctuations in the free-neutron count enhance $\mathrm{Var}\left(\alpha_{\mathrm{ch},\eta}\right)_{\mathrm{mod}}$ and suppress the magnitude of the correlator.

\begin{figure}[!htbp]
\centering
\includegraphics[width=0.45\textwidth]{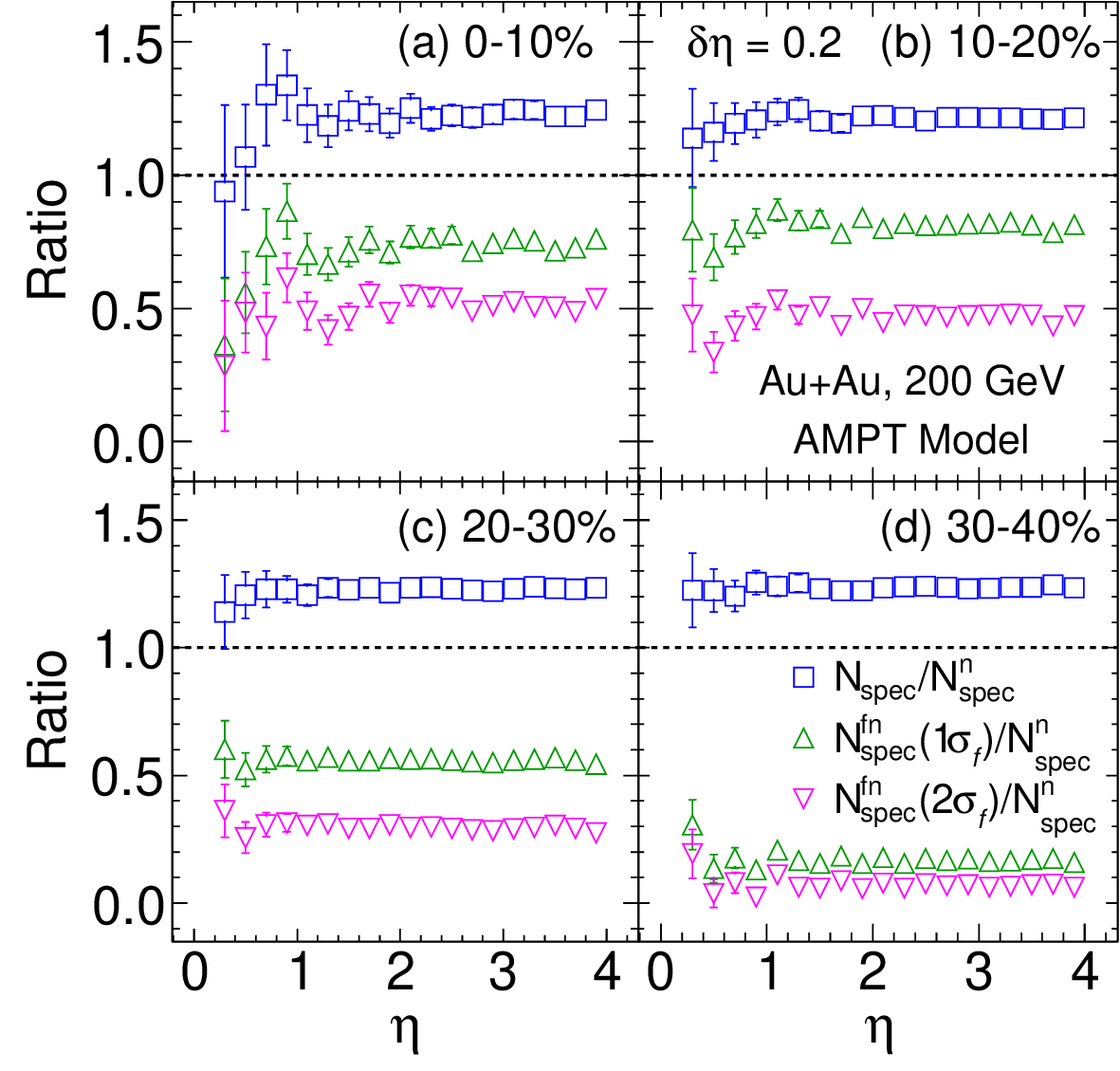}
\caption{Ratio of $\rho(nA_\mathrm{sp},\alpha_{\mathrm{ch},\eta})$ calculated using total spectators ($N_{\mathrm{spec}}$) and free-neutron spectators ($N_{\mathrm{spec}}^{fn}$) relative to that calculated with neutron spectators ($N_{\mathrm{spec}}^{n}$) for different centralities. The $1\sigma_{f}$ and $2\sigma_{f}$ cases represent different levels of fluctuation in $N_{\mathrm{spec}}^{fn}$ due to spectator dynamics.}
\label{fig:sensitivity_ratio}
\end{figure}

Figure~\ref{fig:sensitivity_ratio} quantifies this suppression via the ratio of the correlator calculated with $N_{\mathrm{spec}}$ and $N_{\mathrm{spec}}^{fn}$ relative to that measured with $N_{\mathrm{spec}}^{n}$. The figure illustrates that the modeling used to incorporate spectator fragmentation effects is independent of $\eta$ and serves here as a first-order study. In real data and more advanced models, a clear $\eta$ dependence of these ratios is expected~\cite{Tarafdar:2014oua}.

Furthermore, Fig.~\ref{fig:sensitivity_ratio} displays clearly that the ratio of the correlator calculated with $N_{\mathrm{spec}}$ is approximately 20\% larger than that measured with $N_{\mathrm{spec}}^{n}$. In contrast, the ratio between the correlator measured with $N_{\mathrm{spec}}^{fn}$ and $N_{\mathrm{spec}}^{n}$ decreases from near unity in central collisions to nearly zero in peripheral collisions, with further suppression if spectator-number fluctuations are increased to $2\sigma_f$. This $2\sigma_f$ variation is investigated here to illustrate the correlator’s sensitivity to increased fluctuations in the number of detected spectators, which could arise from an enhancement in spectator fragmentation. These results demonstrate that $\rho(\alpha_{\mathrm{sp}}, \alpha_{\mathrm{ch},\eta})$ serves not only as a probe of preferential emission but also as a sensitive tool for constraining the contribution of spectator breakup to the longitudinal distribution of produced particles in heavy-ion collisions.

\section{Conclusions and Outlook}\label{sec:conclusion}
We have introduced a data-driven observable, $\rho(\alpha_{\mathrm{sp}},\alpha_{\mathrm{ch},\eta})$, that correlates the event-by-event spectator asymmetry with the pseudorapidity-odd component of the final-state charged-particle distribution. By using normalized asymmetries and a bin-width-independent variance, the correlator is robust against detector effects and analysis granularity, making it practical for experimental implementation. Using AMPT-simulated Au+Au collisions at $\sqrt{s_{NN}}=200$~GeV, we demonstrate that $\rho(\alpha_{\mathrm{sp}}, \alpha_{\mathrm{ch},\eta})$ provides a clear, $\eta$-differential measure of preferential emission: it exhibits the expected pseudorapidity-odd shape, vanishes at midrapidity, and increases in magnitude toward forward pseudorapidity.

A study of the correlator components shows that the covariance between $\alpha_{\mathrm{sp}}$ and $\alpha_{\mathrm{ch},\eta}$ is largely centrality-independent, consistent with expectations that the fragmentation function of individual particle-production sources does not vary strongly with collision geometry. The centrality dependence of $\rho(\alpha_{\mathrm{sp}},\alpha_{\mathrm{ch},\eta})$ therefore arises primarily from the interplay between $\mathrm{Var}(\alpha_{\mathrm{sp}})$, which reflects fluctuations in the number of forward- and backward-going participants, and $\mathrm{Var}(\alpha_{\mathrm{ch},\eta})_{\mathrm{mod}}$, which encodes fluctuations in the longitudinal particle-production asymmetry.

A central result of this work is the demonstrated sensitivity of the correlator to spectator dynamics. Incorporating a data-driven model of spectator fragmentation produces a substantial, centrality-dependent suppression of $\rho$. Because spectator breakup modifies the number of free neutrons that reach the ZDCs, the correlator provides an indirect but powerful constraint on the degree of fragmentation, evaporation, and nucleon-nucleon energy exchange in the spectator remnants. Improving constraints on these processes will contribute to a more complete description of the initial-state geometry and the mechanisms by which excited nuclear matter dissociates~\cite{Sorensen:2023zkk,Zheng:2014cha,Shi:2003rmi,NA49:1998ocx}.

The physical motivation underlying this observable parallels the approach developed in Ref.~\cite{Gazdzicki:2005rr}, where correlations between spectator content and produced-particle multiplicities were used to probe longitudinal stopping in fixed-target Pb+Pb collisions at $\sqrt{s_{NN}}=17.3$~GeV. The correlator introduced here provides a 
collider-compatible realization of this idea, formulated entirely in terms of quantities accessible in RHIC experiments and applicable to a broad range of kinematic conditions. In particular, it allows one to investigate how the relationship between initial spectator content and final-state particle production evolves from SPS to RHIC energies, where scenarios closer to transparency dominate.

An experimental program measuring $\rho(\alpha_{\mathrm{sp}},\alpha_{\mathrm{ch},\eta})$ at RHIC and the LHC can significantly sharpen our understanding of the three-dimensional initial state 
while providing new constraints on models of spectator breakup. Extensions to different energies, collision systems, and identified hadrons will help develop a unified, $\eta$-resolved picture of particle production, preferential emission, and spectator dynamics in high-energy nuclear collisions.

\section*{Acknowledgements}
The authors thank Subhasish Chattopadhyay, Jean-Yves Ollitrault, Govert Nijs, Victor Roy, Subhash Singha and Chunjian Zhang for their valuable comments. We also thank Elena Bratkovskaya for her insightful suggestions. SB is supported by a grant of the Dutch Research Council NWO: OCENW.XL21.XL21.038.

\bibliography{LongPartProduction} 

\appendix 
\section{Modified Variance}
\label{Sec:Appendix1}
First, we justify the use of the modified variance $Var(\alpha_{\mathrm{ch},\eta})_{\mathrm{mod}}$ in the proposed correlator. In this study, the distribution for $N_{\mathrm{ch},\eta}$ is calculated using particles within the pseudorapidity range of $\eta \pm \frac{\delta\eta}{2}$. To facilitate subsequent calculations, we express the normalized charged particle asymmetry as:
\begin{equation}
\alpha_\mathrm{ch, \eta} = \frac{N_{\mathrm{ch},\eta} - N_{\mathrm{ch},-\eta}}{N_{\mathrm{ch},\eta} + N_{\mathrm{ch},-\eta}} \equiv \frac{a}{b},
\end{equation}
where $a = N_{\mathrm{ch},\eta} - N_{\mathrm{ch},-\eta}$ and $b = N_{\mathrm{ch},\eta} + N_{\mathrm{ch},-\eta}$.

Next, we perform a Taylor expansion of $\alpha_\mathrm{ch, \eta}$ to first order around the mean of $a$ and $b$, denoted as $\mu_{a}$ and $\mu_{b}$, respectively:
\begin{align}
\alpha_{\mathrm{ch}, \eta} &= \alpha_{\mathrm{ch}, \eta}(\mu_{a}, \mu_{b}) + \left(\frac{\partial \alpha_{\mathrm{ch}, \eta}}{\partial a} \right)\Bigg|_{\mu_{a}, \mu_{b}} \left(a - \mu_{a}\right) \nonumber \\
& \quad + \left(\frac{\partial \alpha_{\mathrm{ch}, \eta}}{\partial b} \right)\Bigg|_{\mu_{a}, \mu_{b}} \left(b - \mu_{b}\right). \nonumber \\
&= \alpha_{\mathrm{ch}, \eta}(\mu_{a}, \mu_{b}) + \frac{1}{\mu_{b}} \left(a - \mu_{a}\right) + \frac{\mu_{a}}{\mu_{b}^{2}} \left(b - \mu_{b}\right).
\end{align}

Now, we can approximate the variance of $\alpha_{\mathrm{ch},\eta}$:
\begin{align}
Var(\alpha_{\mathrm{ch},\eta}) &\approx \left(\frac{\partial \alpha_{\mathrm{ch}, \eta}}{\partial a} \Bigg|_{\mu_{a}, \mu_{b}}\right)^{2} Var(a) + \nonumber \\
&\left(\frac{\partial \alpha_{\mathrm{ch}, \eta}}{\partial b} \Bigg|_{\mu_{a}, \mu_{b}}\right)^{2} Var(b) \nonumber \\
&= \frac{Var(a)}{\mu^{2}_{b}} + \frac{Var(b)\mu^{2}_{a}}{\mu^{4}_{b}}.
\end{align}

In the case of symmetric collisions, and in limit of a large number of events, we have $\mu_{a} = 0$. Assuming $N_{\mathrm{ch},\eta}$ and $N_{\mathrm{ch},-\eta}$ are independent, $Var(a) = Var(b) = 2Var(N_{\mathrm{ch,\eta}})$ and $\mu_{b} = 2\mu(N_{\mathrm{ch,\eta}})$. 
Substituting these relationships yields:
\begin{align}\label{Eq:varmod}
Var(\alpha_{\mathrm{ch},\eta}) &\approx \frac{\sigma^{2}(N_{\mathrm{ch,\eta}})}{2\mu^{2}(N_{\mathrm{ch,\eta}})}.
\end{align}
The expression thus obtained for $Var(\alpha_{\mathrm{ch},\eta})$, defined within an infinitesimally small $\delta\eta$, can be denoted as $\mathcal{V}$. Our objective is to investigate how $\mathcal{V}$ can be derived from measurements conducted within an arbitrarily large $\delta\eta' = k\delta\eta$, where $k > 1$. 

In the limit of local uniformity, where the particle density $d\Nch/d\eta$ varies slowly across the pseudorapidity bin $\delta\eta$, the mean and variance of the particle count are expected to scale linearly with the bin-width $\delta\eta$. We can therefore write:
\begin{align}
Var(N_{\mathrm{ch,\eta}})_{k\delta\eta} &= kVar(N_{\mathrm{ch,\eta}})_{\delta\eta}, \\
\mu(N_{\mathrm{ch,\eta}})_{k\delta\eta} &= k\mu(N_{\mathrm{ch,\eta}})_{\delta\eta}.
\end{align}
Replacing in Eq.~\ref{Eq:varmod}, we obtain,
\begin{align}
    \mathrm{Var}(\alpha_{\mathrm{ch},\eta})_{\delta\eta'} = \mathrm{Var}(\alpha_{\mathrm{ch},\eta})_{\kappa\delta\eta} &\approx \frac{\mathrm{Var}(N_{\mathrm{ch,\eta}})_{\delta\eta}}{2k\mu^2(N_{\mathrm{ch,\eta}})_{\delta\eta}} = \frac{\mathcal{V}}{k} = \frac{\mathcal{V}\delta\eta}{\delta\eta'} \nonumber\\
    \Rightarrow\mathrm{Var}(\alpha_{\mathrm{ch},\eta})_{\delta\eta'} \cdot \delta\eta' &= \mathrm{Var}(\alpha_{\mathrm{ch},\eta}) \cdot \delta\eta
\end{align}
Thus, $Var(\alpha_{\mathrm{ch},\eta})_{\mathrm{mod}} = Var(\alpha_{\mathrm{ch},\eta})_{\delta\eta}.\delta\eta$ is independent of changing $\delta\eta$.

\section{Correction of Unphysical Forward/Backward Asymmetry in AMPT model}
\label{Sec:Appendix2}
Default AMPT (v2.26t9b) exhibits an unphysical forward--backward asymmetry in the single-particle pseudorapidity distribution of symmetric collisions~\cite{ZiWei2022RBRC}, i.e.,
\begin{equation}
\frac{dN_{\rm ch}}{d\eta}(\eta) \neq \frac{dN_{\rm ch}}{d\eta}(-\eta),
\end{equation}
even for symmetric Au+Au collisions. Because the correlator introduced in this work is explicitly sensitive to pseudorapidity-odd structure, this artifact must be removed to avoid contaminating the physics interpretation. To restore the expected symmetry, we apply $\eta$-dependent event weights, computed in each centrality class as
\begin{equation}
w(\eta)=
\frac{
    \tfrac{1}{2}\left[(dN_{\rm ch}/d\eta)(\eta)+(dN_{\rm ch}/d\eta)(-\eta)\right]
     }
     {(dN_{\rm ch}/d\eta)(\eta)}.
\label{eq:weightsApp}
\end{equation}

\begin{figure}[!htbp]
\centering
\includegraphics[width=0.35\textwidth]{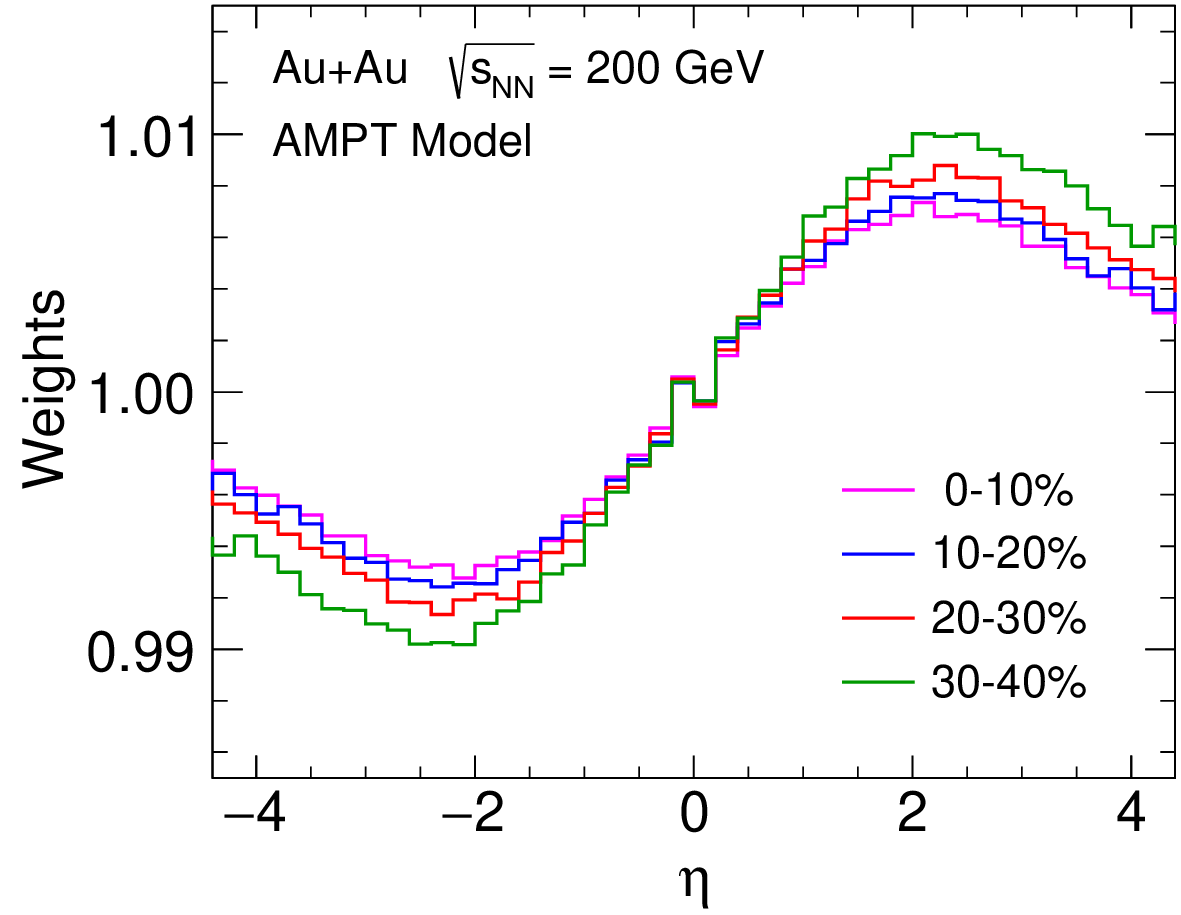}
\caption{Pseudorapidity-dependent weights calculated using Eq.~\ref{eq:weightsApp}, used to symmetrize $dN_{\rm ch}/d\eta$ in the AMPT model.}
\label{fig:weights_symm_app}
\end{figure}
Figure~\ref{fig:weights_symm_app} shows representative weight factors for various centrality intervals, illustrating the magnitude of the correction. 
This procedure replaces the original distribution with its symmetrized form. In each event, the contribution of tracks in a given $\eta$ bin to $dN_{\rm ch}/d\eta$ is multiplied by $w(\eta)$, so that the reweighted distribution in each centrality class becomes
\begin{equation}
\left(\frac{dN_{\rm ch}}{d\eta}\right)_{\mathrm{sym}}(\eta)
= w(\eta)\,\frac{dN_{\rm ch}}{d\eta}(\eta)
= \tfrac{1}{2}\left[\frac{dN_{\rm ch}}{d\eta}(\eta)+\frac{dN_{\rm ch}}{d\eta}(-\eta)\right],
\end{equation}
which is manifestly invariant under $\eta \rightarrow -\eta$. After this reweighting, the resulting charged-particle pseudorapidity density in each centrality interval satisfies
\[
\frac{dN_{\rm ch}}{d\eta}(\eta)=\frac{dN_{\rm ch}}{d\eta}(-\eta),
\]
as required for a symmetric collision system.

\begin{figure}[!htbp]
\centering
\includegraphics[width=0.35\textwidth]{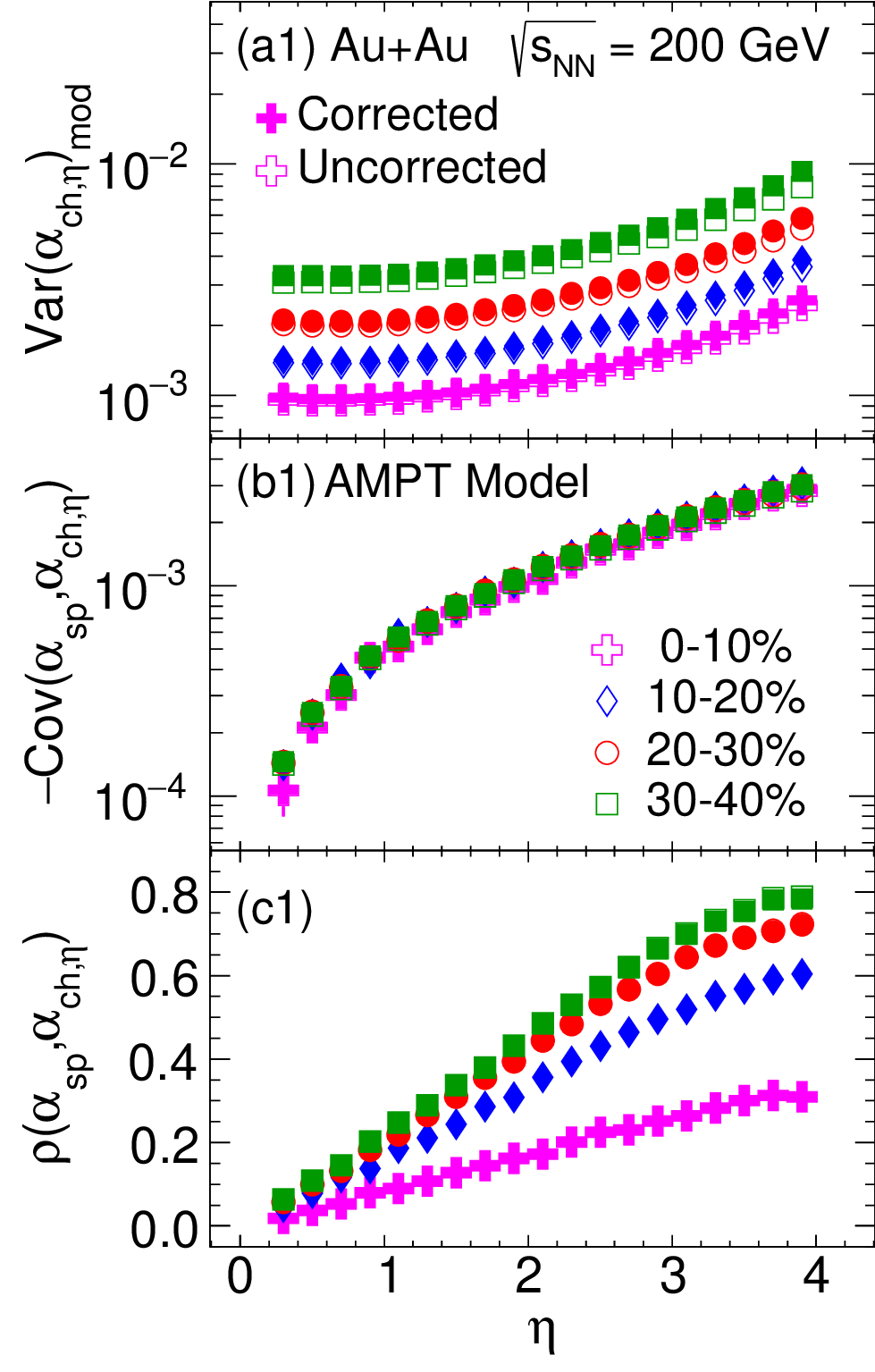}
\caption{Same as Fig.~\ref{fig:main3}, but showing a comparison of (a) $\mathrm{Var}(\alpha_{\mathrm{ch},\eta})_{\mathrm{mod}}$, (b) $-\,\mathrm{Cov}(\alpha_{\mathrm{sp}},\alpha_{\mathrm{ch},\eta})$, and (c) $\rho(\alpha_{\mathrm{sp}},\alpha_{\mathrm{ch},\eta})$ before (open markers) and after (solid markers) symmetrizing $dN_{\rm ch}/d\eta$ in the AMPT model.}
\label{fig:corr_symm_comparison_app}
\end{figure}
The impact of this procedure on the final observables is illustrated in Fig.~\ref{fig:corr_symm_comparison_app}. The uncorrected AMPT sample exhibits a spurious pseudorapidity dependence, most prominently in $\mathrm{Var}(\alpha_{\mathrm{ch},\eta}){\mathrm{mod}}$, which is entirely driven by the unphysical asymmetry in $dN{\rm ch}/d\eta$. In contrast, the correction has only a minor effect on $\rho(\alpha_{\mathrm{sp}},\alpha_{\mathrm{ch},\eta})$. All results presented in this work are therefore obtained using the fully $\eta$-symmetrized event sample.
\end{document}